\documentclass[]{spie}  

\usepackage{amsmath,amsfonts,amssymb}
\usepackage{graphicx}
\usepackage[colorlinks=true, allcolors=blue]{hyperref}
\usepackage{multirow,tabularx}
\usepackage{wrapfig}
\title{Phenotype-preserving metric design for high-content image reconstruction by generative inpainting}

\author[a]{Vaibhav Sharma}
\author[a, b, c]{Artur Yakimovich}
\affil[a]{Center for Advanced Systems Understanding (CASUS), Helmholtz-Zentrum Dresden-Rossendorf e. V. (HZDR), Görlitz, Germany}
\affil[b]{Artificial Intelligence for Life Sciences CIC, Dorset, United Kingdom}
\affil[c]{Institute of Computer Science, University of Wrocław, Wrocław, Poland}

\authorinfo{Correspondence: a.yakimovich@hzdr.de}

\begin{document} 
\maketitle

\begin{abstract}
In the past decades, automated high-content microscopy demonstrated its ability to deliver large quantities of image-based data powering the versatility of phenotypic drug screening and systems biology applications. However, as the sizes of image-based datasets grew, it became infeasible for humans to control, avoid and overcome the presence of imaging and sample preparation artefacts in the images. While novel techniques like machine learning and deep learning may address these shortcomings through generative image inpainting, when applied to sensitive research data this may come at the cost of undesired image manipulation. Undesired manipulation may be caused by phenomena such as neural hallucinations, to which some artificial neural networks are prone. To address this, here we evaluate the state-of-the-art inpainting methods for image restoration in a high-content fluorescence microscopy dataset of cultured cells with labelled nuclei. We show that architectures like DeepFill V2 and Edge Connect can faithfully restore microscopy images upon fine-tuning with relatively little data. Our results demonstrate that the area of the region to be restored is of higher importance than shape. Furthermore, to control for the quality of restoration, we propose a novel phenotype-preserving metric design strategy. In this strategy, the size and count of the restored biological phenotypes like cell nuclei are quantified to penalise undesirable manipulation. We argue that the design principles of our approach may also generalise to other applications.
\end{abstract}


\section{INTRODUCTION}
In the past several decades microscopy took a key role in biomedical discovery and diagnostics \cite{lang2006cellular, niazi2019digital}. Yet, obtaining images of microscopic entities requires sample preparation and staining procedures which are often complex and involve multiple steps \cite{scriven2008image}. This complexity may lead to the occurrence of sample preparation artefacts (SPA) at various stages of the process \cite{lang2006cellular, niazi2019digital}. Some common causes of SPA include incomplete fixation of tissue samples, improper embedding of tissue samples, inefficient removal of dehydration artefacts, inefficient removal of staining artefacts, inclusion of dust particles and excessive tissue damage during sample preparation. The presence of SPAs in micrographs may significantly influence downstream data analysis and interpretation leading to misinterpretation of the biological phenotype. While in some cases these issues can be avoided by improving the handling, as well as imaging of artefact-free samples selectively, in cases of high-content screening (HCS) microscopy SPAs are often inevitable due to the use of automated liquid handling and other high-content sample preparations procedures \cite{boutros2015microscopy, bray2012workflow}.

With the advent of digital image processing, deep learning (DL) and generative models, the removal of SPAs from acquired HCS images may be attained by image inpainting \cite{elharrouss2020image}. In this task, missing or corrupted regions are filled in by a trained generative neural network using context data from the source image itself. However, owing to effects like neural hallucinations \cite{liu2019survey} such image reconstruction tasks may introduce undesired alterations and manipulations of their own. While any generative model manipulates the image to a certain extent, not all alterations are critical. It is therefore crucial to measure how these alterations would affect the perception of the biomedical observation (phenotype) by the image analysis system.

In this work, we employ an open HCI dataset \cite{sharma_vaibhav_2023_1436} containing SPA and assess the ability of the state-of-the-art generative DL algorithms to reconstruct the images through generative inpainting. To ensure the scientific or diagnostic accuracy of the results, we propose a phenotype-preserving metric allowing us to assess the quality of inpainting from the phenotypic point of view. Due to the nature of the dataset, the phenotype-preserving metric we explore in this work focuses on the detection and measurement of cell nuclei. Yet, we argue that similar design principles may be applied to other phenotypes.

\section{RELATED WORK}
Biomedical image reconstruction has been an active area of research for several decades. Introduced in the 1960s, one of the earliest approaches for biomedical image reconstruction was the filtered back projection method \cite{deans2007radon}. This method is widely used in computed tomography and has been improved over the years with variations such as iterative reconstruction and adaptive statistical iterative reconstruction \cite{prakash2010radiation}. Another popular approach is based on the use of compressed sensing techniques. These techniques allow the reconstruction of images from a small number of measurements, which can reduce the amount of data acquisition time required for imaging. Compressed sensing techniques have been applied to various types of biomedical images, including magnetic resonance imaging \cite{lustig2007sparse}, computed tomography \cite{choi2009coded, brady2009compressive}, and positron emission tomography \cite{knoll2014joint}, and microscopy \cite{pavillon2016compressed}. Notably in microscopy, denoising and reconstruction is often associated with enhancing the resolution \cite{gustafsson2008three, muller2016open}.

In addition to these general techniques, there are also specific methods developed for certain types of biomedical images. For example, in ultrasound imaging, time-reversal methods have been used for image reconstruction \cite{lerosey2004time}. In optical coherence tomography, methods based on Fourier domain signal processing have been used \cite{huang1991optical}. However, these techniques are often based on assumptions, which may not hold. Additionally, compressed sensing reconstruction can be computationally expensive, particularly for large volumes of data. To counter the above limitations, DL approaches have also been used for biomedical image reconstruction. Specifically, convolutional neural networks (CNNs) have been used for image superresolution, denoising, and artefact reduction \cite{weigert2018content, qiao2021evaluation, dong2015image}.

However, the assessment of the quality of biomedical image reconstruction is currently limited to well-established computer vision metrics, which may not capture the intricacy required for scientific and diagnostic accuracy. Metrics such as peak signal-to-noise ratio (PSNR) \cite{elbadawy1998information} and structural similarity index measure (SSIM) \cite{wang2004image} are often used to assess quality, yet are known to lack complexity \cite{ponomarenko2007between}. One approach for circumventing this is to evaluate the metric on multiple scales \cite{wang2003multiscale}. Other approaches aim to introduce subjective quality assessment involving human judgement \cite{farrell1999image}. In this work, we propose a novel metric that incorporates both subjective and objective quality assessment of HCI microscopy image reconstruction. The main criterion of the metric we propose is to preserve the biological phenotype from the perspective of image quantification. We argue that a conceptually similar metric design can be proposed for other biomedical image modalities.

\section{METHODS}
\subsection{Code Availability}
The Python source code developed in this work is available under GPLv3 open-source license at \\ \hyperlink{https://github.com/casus/PhIRM}{github.com/casus/PhIRM}.
\subsection{Computational Setup}
All the inpainting models used in this work have been trained using high-performance computing. The Hemera HPC system has been used for this purpose. All the experiments have been carried out using a single 32 GB NVIDIA V100 GPU with 8 (eight) CPU cores. The maximum memory per CPU was set to 1443 MB.

\subsection{Sample preparation Artefacts Masks Generation}
The multispectral properties of the dataset presented \cite{sharma_vaibhav_2023_1436} in this work have been used to generate ground truth images containing exclusively nuclei or SPA. The current methodology involves the generation of artefact masks from 2160x2160 pixel TIFF images obtained from the CFP channel. Otsu thresholding \cite{otsu1979threshold} has been implemented to obtain a threshold value, which is subsequently multiplied by 0.7 to enhance the quality of the masks. We observed that Otsu's thresholding often eliminates small artefacts or boundary pixels of large artefacts, and hence the multiplication factor was introduced. The resultant image is binarised and subjected to morphological operations in the form of opening and closing.

\subsection{Artificial Masks Generation}
To understand the influence of mask shape on inpainting performance we have created two kinds of artificial masks: rectangular masks and irregular masks. In the case of the first, we created rectangular mask datasets which contain images with varying mask area: 10-20\%, 20-30\%, 30-40\% and 40-50\%. For example, A 10-20\% rectangular mask dataset contains images having 10-20\% of the total image area covered by a rectangular mask. This is done by placing white rectangles of a particular area range (e.g.: 10-20\% of the entire image area) at random regions inside a black image of size 256x256.

To train models with irregular-shaped masks of random sizes, we used an approach to draw masks based on randomly positioned lines with a predefined minimum and maximum number of vertices. Our approach rotates angles and produces thick lines by joining these vertices. Then it puts circles in the intersections of the two lines to guarantee their smoothness. This algorithm has been adapted from the DeepFillV2 architecture \cite{yu2019free}.

\subsection{Data Augmentation}
To increase the size of our dataset we have developed an image patch (zoomed-image) generator that extracts smaller zoomed and cropped images from a larger image of size 2160x2160 pixels. The generator generates patches of size 256x256, starting from the left-hand side of the image and moving towards the right until the entire image is covered. After normalization using the min-max method, the resulting intensity range is 0 to 255. All 256x256-sized nuclei and mask images used in this work have been created using this zoomed image generator.

\subsection{Model Fine-tuning}
To ensure good performance of the Context Encoder, DeepFill V2 and Edge Connect pre-trained networks, these models were fine-tuned using microscopy data. For this we have generated and augmented dataset containing 36112 data points with masks constructed as described in respective sections above. Context Encoder was fine-tuned for 200 epochs with learning rate 0.0002. DeepFill V2 was fine-tuned for 20 epochs with learning rate 0.0001. Edge Connect was fine-tuned for 68 epochs with learning rate 0.0001. All networks were trained until convergence.

\subsection{Phenotype-preserving Image Reconstruction Metric Design}
In the case of nuclear images, PhIRM can be designed to contain the following components: nuclear count difference defined in Equation~(\ref{eq:ncd}), nuclear area difference defined in Equation~(\ref{eq:nad}) and artefact area difference defined in Equation~(\ref{eq:aad}). These values can be computed as follows:

\begin{equation}
\label{eq:ncd}
NCD = 
    \begin{cases}
        0 \cdot 0 & \text if \alpha = 0\\
        1 \cdot 1^\alpha & \text if \alpha >0\\
        2^{|\alpha|} & \text if \alpha <0,\\
    \end{cases}
\end{equation}

where $\alpha$ (default value is: 1.1) is the difference in number of nuclei between the original and reconstructed image.

\begin{equation}
\label{eq:nad}
NAD = \omega_{NAD} \cdot (A_{nuc \ out} - A_{nuc \ in}),
\end{equation}

where $A_{nuc \ out}, A_{nuc \ in}$ is the area of nuclei in the original and reconstructed image, and $\omega_{NAD}$ (default value is: 0.0002) is the respective weight.

\begin{equation}
\label{eq:aad}
AAD = \omega_{AAD} \cdot (A_{art \ out} - A_{art \ in}),
\end{equation}

where $A_{art \ out}, A_{art \ in}$ is the area of artefacts in the original and reconstructed image, and $\omega_{AAD}$ (default value is: 0.001) is the respective weight. These components can be computed in the following algorithm.

\noindent\textbf{Algorithm 1. PhIRM Factors Calculation}\\
1. Apply Otsu thresholding \cite{otsu1979threshold} to the input image to produce a binary mask image.\\
2. Perform connected components analysis on the binary image.\\
3. Avoid components with an area of less than 50 px.\\
4. Compute of mean and maximum values for each component.\\
5. Identify components as artefacts if the maximum value computed in step 4 equals 255 and the mean value is greater than or equal to 210.\\ 
    Otherwise identification as nuclei.\\
6. Identify components as single nuclei if the maximum value computed in step 4 is less than 255 and the area of the component is less than 2200. \\
    Otherwise, identify as a patch containing two overlapping nuclei.\\
7. Store the total number of nuclei, total nuclei area, and total artefact area.

After these steps are applied to both the source and inpainted image, the difference in the number of nuclei, total nuclei area, and total artefact area between the two images can be computed. These attributes are then used to calculate the final score defined in the following Equation~(\ref{eq:PhIRM}).

\begin{equation}
\label{eq:PhIRM}
PhIRM = \frac{10 - (NCD + NAD + AAD)}{10}.
\end{equation}
\section{RESULTS}

\begin{figure}[ht]
\includegraphics[width=1\linewidth]{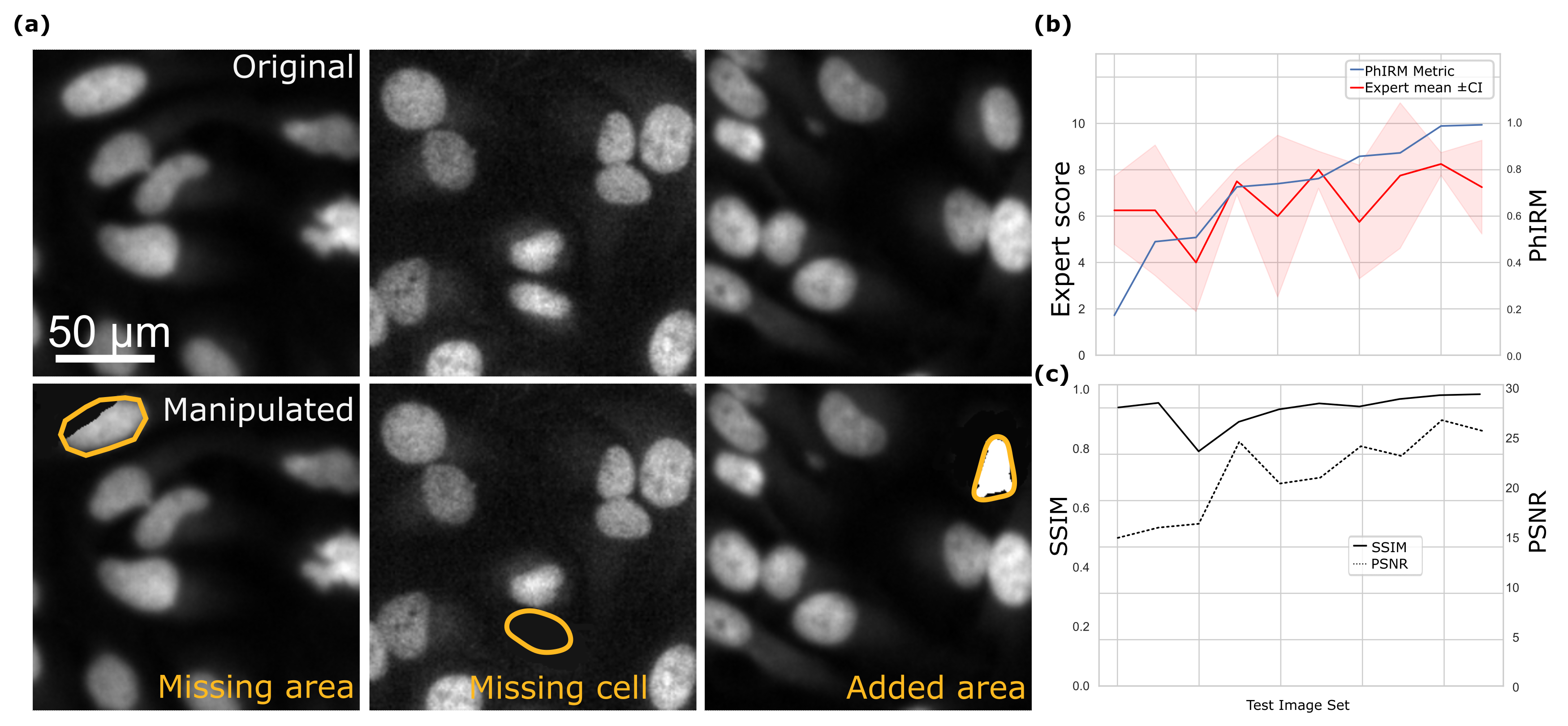}
\caption{ \label{fig:fig1} 
\textbf{Metric validation set for comparison to expert opinion.} (a) examples of images in the validation set with respective alterations. (b) Comparison of the expert opinion to the PhIRM metric measured on this test image set. (c) Comparison of peak signal-to-noise ratio (PSNR) and structural similarity index metric (SSIM) on the same set.}
\end{figure} 

\subsection{Metric design and validation}
HCI microscopy containing SPA may be corrected computationally using generative inpainting. However, the assessment of the scientific accuracy of such corrected images is problematic. To address that we proposed Phenotype-preserving Image Reconstruction Metric (PhIRM). At its core, PhIRM is assessing whether the phenotype-relevant information is still preserved upon image inpainting. In the case of high-content fluorescence microscopy of cell nuclei imaged in the presence or absence of SPA \cite{sharma_vaibhav_2023_1436}, the preservation of the phenotype constitutes factors including the area of the nuclei, their count, as well as the area of the remaining artefact. These factors were taken into account and heuristically weighted for this specific phenotype and task (see Methods).

Next, to assess how PhIRM compares to expert assessment we have constructed a test set containing the original image together with images manipulated in a manner to corrupt the phenotype. Specifically, manipulations created images with the missing nuclear area, missing nuclei or introduced artefacts (Fig \ref{fig:fig1}a). We then asked four image analysis experts (including one senior author of this work) to assess the difference between the original and manipulated images from the perspective of phenotype consistency. Remarkably, despite the judgement discrepancies in the extreme cases, overall expert opinion showed a good agreement with PhIRM (Fig \ref{fig:fig1}b).

To understand how PhIRM would compare to existing image reconstruction metrics like SSIM and PSNR we measured these on the same set of images. (Fig \ref{fig:fig1}c). Our comparison suggested that while PSNR somewhat correlated with the expert opinion, SSIM completely failed to capture the differences. Notably, in extreme cases like very low consistency or very high consistency, PhIRM seems to capture the differences better than PSNR. Interestingly, these were the cases experts had a rather low consensus on as well. We attributed these observations to an eventuality that both experts and PSNR would likely penalise differences that may not be crucial for phenotypic measurements. PhIRM at the same time would focus exclusively on aspects decisive for the phenotypic accuracy.

\subsection{High-content image reconstruction by generative inpainting}
\begin{figure}[ht]
\includegraphics[width=1\linewidth]{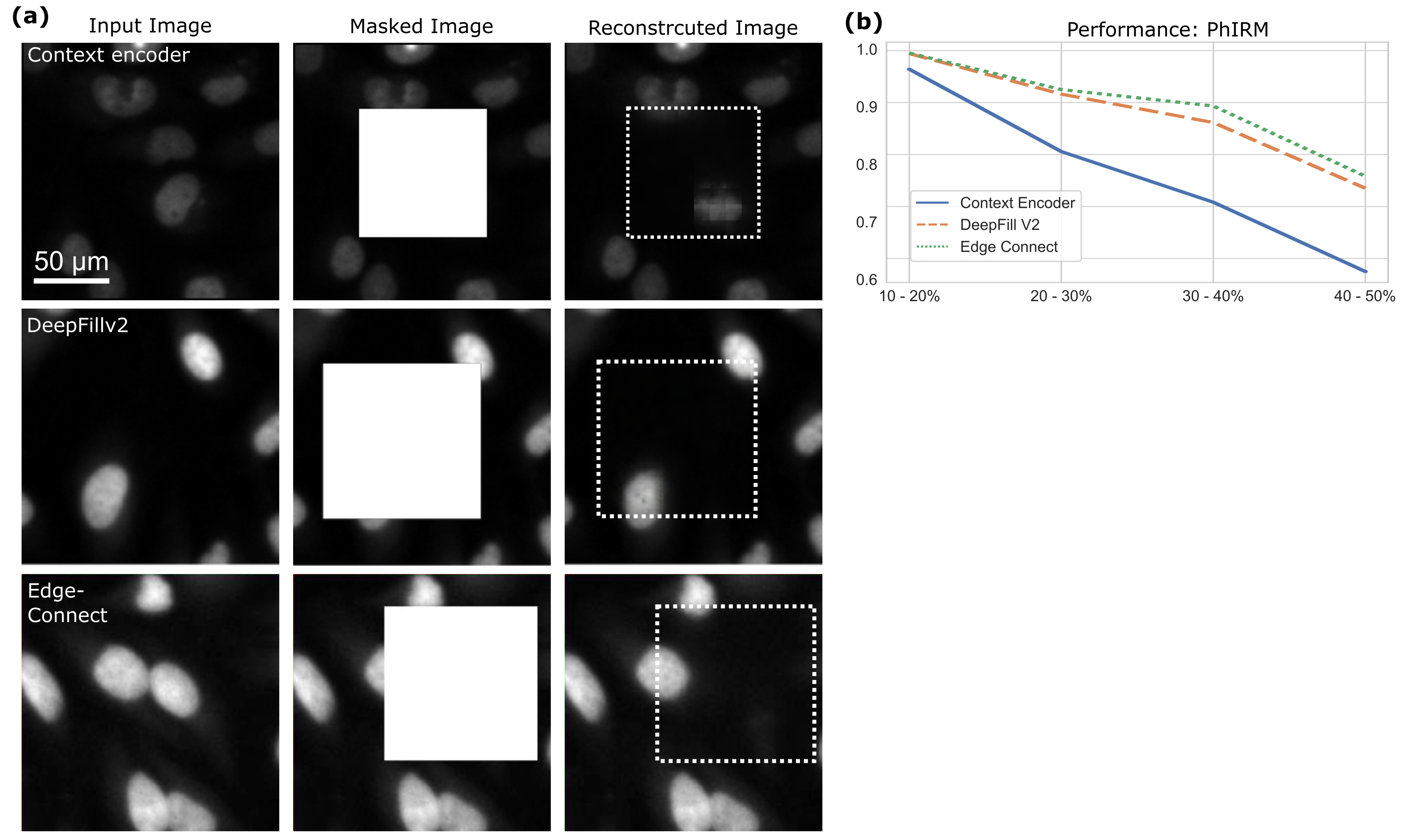}
\caption{\label{fig:fig2} 
\textbf{State-of-the-art image inpainting architectures performance on high-content microscopy} (a) example of images with square masks covering 30-40\% of the area with respective inpainting results using pre-trained Context Encoder, DeepFill V2 and Edge Connect networks. (b) PhIRM metric output for Context Encoder, DeepFill V2 and Edge Connect networks depending on the area covered by the mask.}
\end{figure}

\begin{figure}[ht]
\includegraphics[width=1\linewidth]{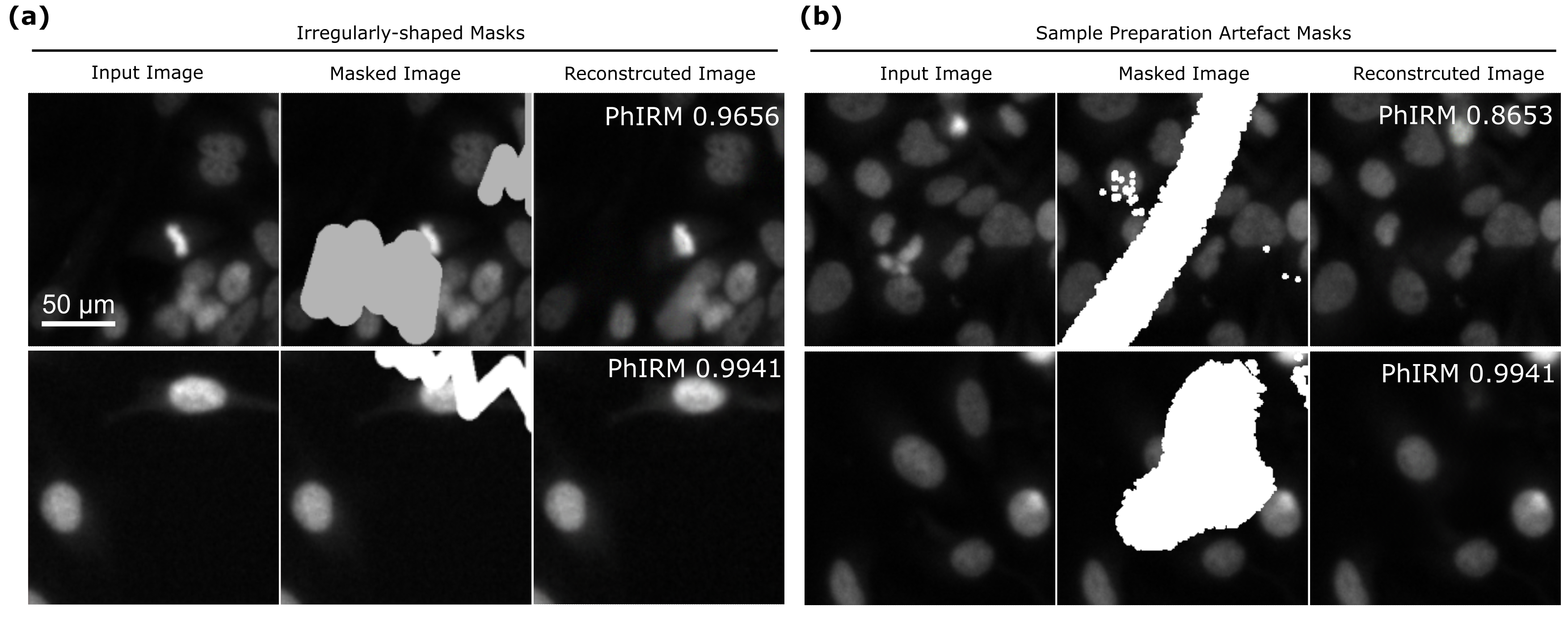}
\caption{\label{fig:fig3} 
\textbf{Edge Connect architecture performance on irregularly-shaped masks and sample preparation artefacts.} (a) Examples of Edge Connect reconstruction on irregularly-shaped masks, accompanied by phenotype-preserving image reconstruction metric (PhIRM). (b) Examples of Edge Connect reconstruction on sample preparation artefact masks, accompanied by PhIRM.}
\end{figure}

To measure the ability of generative inpainting to restore high-content images we selected three state-of-the-art pre-trained models with varying expressive capacity of the deep artificial neural networks (DNN) they are based on. Namely, we selected Context Encoder \cite{pathak2016context}, DeepFill V2 \cite{yu2019free} and Edge Connect \cite{nazeri2019edgeconnect}. Next, we fine-tuned these models on our dataset and evaluated their performance using PhIRM. It is worth noting, however, that Context Encoder was not designed to perform inpainting with irregularly shaped masks. Therefore, to ensure a fair comparison, we have first created a synthetic set of images with rectangular masks (Fig \ref{fig:fig2}a). To understand how the area of the mask impacts performance we have used four distinct ranges 10-20\%, 20-30\%, 30-40\%, 40-50\% (Fig \ref{fig:fig2}b). Our comparison showed that Deep Fill V2 and Edge Connect performed significantly better than the Context Encoder, especially with the largest masks. On mask area above 30\% Edge Connect outperformed Deep Fill V2.

Finally, we compared the performance of Edge Connect architecture on rectangular masks to the performance on irregularly-shaped masks (Fig \ref{fig:fig3}a) and the masks derived from high-content image SPAs (Fig \ref{fig:fig3}b). Remarkably, despite quite obvious differences in shapes, with PhIRM between 0.86 and 0.99 the performance remained high for both types of masks. Similarly to the case of rectangular masks, performance seemed to be stronger connected to the area obstructed, rather then shape of the mask. Based on this we concluded that the state-of-the-art Edge Connect architecture may be employed for reconstructive inpainting of high-content fluorescent microscopy images obstructed by SPAs without significant alteration of nuclear phenotype.

\section{CONCLUSIONS}
This research highlights the potential of using generative inpainting for microscopy image reconstruction. While generative inpainting bears a great promise to microscopy reconstruction, it inevitably creates artefacts. To ensure the scientific accuracy of such reconstructed images we propose a novel metric - phenotype-preserving image reconstruction metric (PhIRM). The metric score ranges from 0.0 to 1.0, with higher scores indicating better inpainting quality. Compared to traditional metrics like PSNR and SSIM, the proposed metric demonstrates higher relevance and sensitivity. PhIRM aims to take into account only the changes crucial for the preservation of the biological phenotype. In the specific case of quantification of fluorescent cell nuclei in the presence of sample preparation artefacts in high-content imaging, these changes are cell count, cell area and artefact area. We demonstrate that the metric we proposed is in good agreement with expert opinion.

Noteworthy, in order to make PhIRM useful for other applications, a different set of phenotype factors may need to be constructed. While this may seem to limit the application of PhIRM in an off-the-shelf manner, we argue that the construction of PhIRM for each individual application should be rather straightforward for a subject matter expert. For example, should cell cycle detection be important for the phenotype, one could include the difference in the total fluorescence intensity of the marker. 

Furthermore, we employed the PhIRM metric to evaluate the state-of-the-art inpainting architectures including Context Encoder\cite{pathak2016context}, DeepFill V2 \cite{yu2019free} and Edge Connect \cite{nazeri2019edgeconnect}. We demonstrate that both visually and according to the PhIRM values DeepFill V2 and Edge Connect outperformed an older Context Encoder architecture. Additionally, we showed that for our particular task Edge Connect performed consistently better. Remarkably, we showed that, in the case of Edge Connect inpainting, the size of the damaged area significantly affects the inpainting quality. However, the shape of the damaged regions had only a minor impact. In summary, in this work, we argue that if accompanied by an adequate metric generative inpainting may be useful for image reconstruction in HCI microscopy.

\acknowledgments 
We thank Dr. Vardan Andriasyan and Anthony Petkidis for their expert opinion on microscopy image evaluation. This work was partially funded by the Center for Advanced Systems Understanding (CASUS) which is financed by Germany’s Federal Ministry of Education and Research (BMBF) and by the Saxon Ministry for Science, Culture, and Tourism (SMWK) with tax funds on the basis of the budget approved by the Saxon State Parliament. 

\bibliography{report} 

\begin{thebibliography}{10}

\bibitem{lang2006cellular}
Lang, P., Yeow, K., Nichols, A., and Scheer, A., ``Cellular imaging in drug
  discovery,'' {\em Nature Reviews Drug Discovery}~{\bf 5}(4),  343--356
  (2006).

\bibitem{niazi2019digital}
Niazi, M. K.~K., Parwani, A.~V., and Gurcan, M.~N., ``Digital pathology and
  artificial intelligence,'' {\em The lancet oncology}~{\bf 20}(5),  e253--e261
  (2019).

\bibitem{scriven2008image}
Scriven, D.~R., Lynch, R.~M., and Moore, E.~D., ``Image acquisition for
  colocalization using optical microscopy,'' {\em American Journal of
  Physiology-Cell Physiology}~{\bf 294}(5),  C1119--C1122 (2008).

\bibitem{boutros2015microscopy}
Boutros, M., Heigwer, F., and Laufer, C., ``Microscopy-based high-content
  screening,'' {\em Cell}~{\bf 163}(6),  1314--1325 (2015).

\bibitem{bray2012workflow}
Bray, M.-A., Fraser, A.~N., Hasaka, T.~P., and Carpenter, A.~E., ``Workflow and
  metrics for image quality control in large-scale high-content screens,'' {\em
  Journal of biomolecular screening}~{\bf 17}(2),  266--274 (2012).

\bibitem{elharrouss2020image}
Elharrouss, O., Almaadeed, N., Al-Maadeed, S., and Akbari, Y., ``Image
  inpainting: A review,'' {\em Neural Processing Letters}~{\bf 51},  2007--2028
  (2020).

\bibitem{liu2019survey}
Liu, H., Zheng, X., Han, J., Chu, Y., and Tao, T., ``Survey on gan-based face
  hallucination with its model development,'' {\em IET Image Processing}~{\bf
  13}(14),  2662--2672 (2019).

\bibitem{sharma_vaibhav_2023_1436}
Sharma, V. and Yakimovich, A., ``{High-content multi-spectral fluorescence
  microscopy sample preparation artefacts},'' (Jan. 2023).

\bibitem{deans2007radon}
Deans, S.~R.,  [{\em The Radon transform and some of its
  applications}{\nolinebreak\hspace{0.1em}]}, Courier Corporation (2007).

\bibitem{prakash2010radiation}
Prakash, P., Kalra, M.~K., Digumarthy, S.~R., Hsieh, J., Pien, H., Singh, S.,
  Gilman, M.~D., and Shepard, J.-A.~O., ``Radiation dose reduction with chest
  computed tomography using adaptive statistical iterative reconstruction
  technique: initial experience,'' {\em Journal of computer assisted
  tomography}~{\bf 34}(1),  40--45 (2010).

\bibitem{lustig2007sparse}
Lustig, M., Donoho, D., and Pauly, J.~M., ``Sparse mri: The application of
  compressed sensing for rapid mr imaging,'' {\em Magnetic Resonance in
  Medicine: An Official Journal of the International Society for Magnetic
  Resonance in Medicine}~{\bf 58}(6),  1182--1195 (2007).

\bibitem{choi2009coded}
Choi, K. and Brady, D.~J., ``Coded aperture computed tomography,'' in [{\em
  Adaptive Coded Aperture Imaging, Non-Imaging, and Unconventional Imaging
  Sensor Systems}{\nolinebreak\hspace{0.1em}]},   {\bf 7468},  99--108, SPIE
  (2009).

\bibitem{brady2009compressive}
Brady, D.~J., Choi, K., Marks, D.~L., Horisaki, R., and Lim, S., ``Compressive
  holography,'' {\em Optics express}~{\bf 17}(15),  13040--13049 (2009).

\bibitem{knoll2014joint}
Knoll, F., Koesters, T., Otazo, R., Block, T., Feng, L., Vunckx, K., Faul, D.,
  Nuyts, J., Boada, F., and Sodickson, D.~K., ``Joint reconstruction of
  simultaneously acquired mr-pet data with multi sensor compressed sensing
  based on a joint sparsity constraint,'' {\em EJNMMI physics}~{\bf 1},  1--2
  (2014).

\bibitem{pavillon2016compressed}
Pavillon, N. and Smith, N.~I., ``Compressed sensing laser scanning
  microscopy,'' {\em Optics express}~{\bf 24}(26),  30038--30052 (2016).

\bibitem{gustafsson2008three}
Gustafsson, M.~G., Shao, L., Carlton, P.~M., Wang, C.~R., Golubovskaya, I.~N.,
  Cande, W.~Z., Agard, D.~A., and Sedat, J.~W., ``Three-dimensional resolution
  doubling in wide-field fluorescence microscopy by structured illumination,''
  {\em Biophysical journal}~{\bf 94}(12),  4957--4970 (2008).

\bibitem{muller2016open}
M{\"u}ller, M., M{\"o}nkem{\"o}ller, V., Hennig, S., H{\"u}bner, W., and Huser,
  T., ``Open-source image reconstruction of super-resolution structured
  illumination microscopy data in imagej,'' {\em Nature communications}~{\bf
  7}(1),  10980 (2016).

\bibitem{lerosey2004time}
Lerosey, G., De~Rosny, J., Tourin, A., Derode, A., Montaldo, G., and Fink, M.,
  ``Time reversal of electromagnetic waves,'' {\em Physical review
  letters}~{\bf 92}(19),  193904 (2004).

\bibitem{huang1991optical}
Huang, D., Swanson, E.~A., Lin, C.~P., Schuman, J.~S., Stinson, W.~G., Chang,
  W., Hee, M.~R., Flotte, T., Gregory, K., Puliafito, C.~A., et~al., ``Optical
  coherence tomography,'' {\em science}~{\bf 254}(5035),  1178--1181 (1991).

\bibitem{weigert2018content}
Weigert, M., Schmidt, U., Boothe, T., M{\"u}ller, A., Dibrov, A., Jain, A.,
  Wilhelm, B., Schmidt, D., Broaddus, C., Culley, S., et~al., ``Content-aware
  image restoration: pushing the limits of fluorescence microscopy,'' {\em
  Nature methods}~{\bf 15}(12),  1090--1097 (2018).

\bibitem{qiao2021evaluation}
Qiao, C., Li, D., Guo, Y., Liu, C., Jiang, T., Dai, Q., and Li, D.,
  ``Evaluation and development of deep neural networks for image
  super-resolution in optical microscopy,'' {\em Nature Methods}~{\bf 18}(2),
  194--202 (2021).

\bibitem{dong2015image}
Dong, C., Loy, C.~C., He, K., and Tang, X., ``Image super-resolution using deep
  convolutional networks,'' {\em IEEE transactions on pattern analysis and
  machine intelligence}~{\bf 38}(2),  295--307 (2015).

\bibitem{elbadawy1998information}
Elbadawy, O., El-Sakka, M.~R., and Kamel, M.~S., ``An information theoretic
  image-quality measure,'' in [{\em Conference Proceedings. IEEE Canadian
  Conference on Electrical and Computer Engineering (Cat. No.
  98TH8341)}{\nolinebreak\hspace{0.1em}]},   {\bf 1},  169--172, IEEE (1998).

\bibitem{wang2004image}
Wang, Z., Bovik, A.~C., Sheikh, H.~R., and Simoncelli, E.~P., ``Image quality
  assessment: from error visibility to structural similarity,'' {\em IEEE
  transactions on image processing}~{\bf 13}(4),  600--612 (2004).

\bibitem{ponomarenko2007between}
Ponomarenko, N., Silvestri, F., Egiazarian, K., Carli, M., Astola, J., and
  Lukin, V., ``On between-coefficient contrast masking of dct basis
  functions,'' in [{\em Proceedings of the third international workshop on
  video processing and quality metrics}{\nolinebreak\hspace{0.1em}]},   {\bf
  4}, Scottsdale USA (2007).

\bibitem{wang2003multiscale}
Wang, Z., Simoncelli, E.~P., and Bovik, A.~C., ``Multiscale structural
  similarity for image quality assessment,'' in [{\em The Thrity-Seventh
  Asilomar Conference on Signals, Systems \& Computers,
  2003}{\nolinebreak\hspace{0.1em}]},   {\bf 2},  1398--1402, Ieee (2003).

\bibitem{farrell1999image}
Farrell, J.~E., ``Image quality evaluation,'' {\em Colour imaging: vision and
  technology} ,  285--313 (1999).

\bibitem{otsu1979threshold}
Otsu, N., ``A threshold selection method from gray-level histograms,'' {\em
  IEEE transactions on systems, man, and cybernetics}~{\bf 9}(1),  62--66
  (1979).

\bibitem{yu2019free}
Yu, J., Lin, Z., Yang, J., Shen, X., Lu, X., and Huang, T.~S., ``Free-form
  image inpainting with gated convolution,'' in [{\em Proceedings of the
  IEEE/CVF international conference on computer
  vision}{\nolinebreak\hspace{0.1em}]},   4471--4480 (2019).

\bibitem{pathak2016context}
Pathak, D., Krahenbuhl, P., Donahue, J., Darrell, T., and Efros, A.~A.,
  ``Context encoders: Feature learning by inpainting,'' in [{\em Proceedings of
  the IEEE conference on computer vision and pattern
  recognition}{\nolinebreak\hspace{0.1em}]},   2536--2544 (2016).

\bibitem{nazeri2019edgeconnect}
Nazeri, K., Ng, E., Joseph, T., Qureshi, F.~Z., and Ebrahimi, M.,
  ``Edgeconnect: Generative image inpainting with adversarial edge learning,''
  {\em arXiv preprint arXiv:1901.00212}  (2019).

\end{thebibliography}
\bibliographystyle{spiebib} 

\end{document}